\documentclass{JHEP}
\preprint{ {\tt hep-th/0005085} }
\newcommand{\be}{\begin{equation}}
\newcommand{\ee}{\end{equation}}
\newcommand{\ba}{\begin{array}{l}}
\newcommand{\ea}{\end{array}}
\newcommand{\bea}{\begin{eqnarray}}
\newcommand{\eea}{\end{eqnarray}}
\newcommand{\eq}[1]{(\ref{#1})}

\newcommand{\aq}{ A^{(2)} }
\newcommand{\ac}{ A^{(3)} }
\newcommand{\taq}{ \tilde{A}^{(2)} }
\newcommand{\tac}{ \tilde{A}^{(3)} }
\newcommand{\ta}{ \tilde{A} }
\newcommand{\tphi}{\tilde{\phi}}
\newcommand{\el}{E^{(0)}}
\newcommand{\es}{E^{(1)}}
\newcommand{\cl}{C^{(0)}}
\newcommand{\cq}{C^{(1)}}
\newcommand{\dl}{D^{(0)}}
\newcommand{\dq}{D^{(1)}}
\newcommand{\kl}{K^{(1)}}
\newcommand{\bl}{B^{(2)}}
\newcommand{\ph}[2]{\phi_{#1}^{(#2)} }
\newcommand{\et}[2]{\eta_{#1}^{(#2)} }
\newcommand{\hph}[2]{ \hat{\phi}_{#1}^{(#2)} }
\newcommand{\bph}[2]{ \bar{\phi}_{#1}^{(#2)} }
\newcommand{\het}[2]{ \hat{\eta}_{#1}^{(#2)} }
\newcommand{\bet}[2]{ \bar{\eta}_{#1}^{(#2)} }
\title{$U(1)$ gauge invariance from open string field theory}
\author{Justin R. David\\ Department of Physics\\University of
California\\Santa Barbara, CA 93106\\USA.\\
\email{justin@vulcan.physics.ucsb.edu} }
\abstract{ The naive low energy effective action of the tachyon and
the $U(1)$ gauge field obtained from string field theory does not
correspond to the world volume action of unstable branes in
bosonic string theory. 
We show that there exists a
field redefinition which relates 
the gauge field and the tachyon  of the string
field theory action 
to the fields in the world volume action of
unstable branes.
We identify a string gauge symmetry
which corresponds to the $U(1)$ gauge transformation. This 
is done to the
first non linear order in the fields.  
We examine the vector fluctuations at the tachyon condensate 
till level (4,8). }

\begin{document}
\baselineskip 4.0ex

\section{Introduction}

Recent works have shown that
string field theory is   useful to understand the physics of tachyon
condensation \cite{sen}. 
In bosonic string theory, the decay of the D 25-brane
has been shown to correspond to the rolling of the tachyon of open
bosonic string theory down to a new stationary point in the tachyon
potential. The analysis of the tachyon potential was carried out using
string field theory. It was also seen that level truncation is a good
approximation scheme for the calculation 
of the value of the minimum of the tachyon potential and the physics
of solitons in the potential
\cite{kossam,senzwi,wati,moewati,harkra,jevtar,moesenzwi}. 
The level truncation
seems to be a good approximation scheme in superstring field theory
\cite{ber,senzwiber,smerae,iqbnaq}.  

As the D 25-brane has completely decayed at the new stationary point of
the tachyon potential, there should be no dynamical open string
degrees of freedom. In particular the $U(1)$ gauge field should no
longer be dynamical. 
It was conjectured in \cite{sen} that at this stationary point the
effective action does not have a kinetic term for the $U(1)$
gauge field. In fact the term multiplying the kinetic term was argued
to be the tachyon potential \cite{sen2}.
This would render this field auxiliary. It was also
suggested it would be interesting to see this phenomenon directly from
string field theory.
The difficulty with the direct evaluation of the 
coefficient of the kinetic term of the
gauge field using string field theory is the problem of field
redefinition. 

We will now state this problem.
Consider eliminating all the massive modes and the auxiliary fields of
the string field using the equations of motion.  
We obtain the effective action ${\cal S}
(\ta_\mu, \tilde{\phi})$ as a function of the $U(1)$ gauge field and
the tachyon $\tilde{\phi}$ at the tree level. 
Can we identify the gauge field $\ta_\mu$ and the tachyon
$\tilde{\phi}$ with that of the gauge field of the D25-brane and its
tachyon, and thus obtain the effective action of the D25-brane from
string field theory ?
A string field $\Phi$ has the following gauge transformation in string
field theory
\be
\delta{\Phi} = Q_{B} \Lambda +  \Lambda * \Psi - \Psi *\Lambda
\ee
Here $\Lambda$, the infinitesimal gauge parameter
is a string field of ghost number zero and `$*$' is the
star product in string field theory. Let us call the  $U(1)$ 
field present in the expansion of $\Phi$ as $\tilde{A}_\mu$. Then, it
is clear that $\tilde{A}_\mu$ does not transform as 
\be
\delta\tilde{A}_\mu = \partial_\mu\epsilon
\ee
The transformation property of $\tilde{A}$ has extra terms arising 
from the non-Abelian like transformation property of the string field
$\Phi$. Therefore the gauge field that appears in the low energy
effective Lagrangian $A_\mu$ which has the conventional gauge
transformation property cannot be identified with $\tilde{A}_\mu$. 
Thus the naive low energy effective action ${\cal S} (\ta_\mu,
\tilde{\phi} )$ cannot be identified with the world volume action 
of the unstable brane.

It must be possible to find a relationship between $A_\mu$ and
$\tilde{A}_{\mu}$ and $\phi$ and $\tilde{\phi}$.
This will involve  a field redefinition 
\be
\label{redef}
\ta_\mu = \ta_\mu (A_\mu, \phi), \;\;\; \tilde{\phi} =
\tilde{\phi}(A_\mu, \phi)
\ee
It also involves the identification of the string field $\Lambda$
which corresponds to the conventional gauge transformation property of
$A_\mu$. 
In this paper we  show that it is indeed possible to find such a
relationship at the first non-linear level in the fields. To do this
we use methods developed to identify general coordinate
transformation and anti-symmetric tensor gauge transformation as a set
of string field theoretic symmetries developed for the non-polynomial
closed string field theory by \cite{ghosen}\footnote{The author thanks
A. Sen for pointing out this reference.}. 

The fact that gauge field of the low energy effective action involves
not only $\tilde{A}_\mu$ but also other fields is important in
determining the coefficient of the kinetic term of $A_\mu$ from string
field theory. Determining the naive kinetic term of $\tilde{A}_\mu$
is not enough. This field redefinition involved in relating $A_\mu$
and $\tilde{A}_\mu$ is also important in determining terms of the
non-Abelian Dirac Born Infeld action from string field theory as done
in \cite{wati}. This would help in understanding the  
discrepancy in gauge invariance
which arose in trying to determining the coefficients of
certain terms in the non-Abelian Dirac Born Infeld action from string
field theory.

Thus it seems difficult to compare the effective action ${\cal S}
(\ta_\mu, \tilde{\phi})$ obtained from string field theory 
with the world volume action of unstable
branes and verify the claims of \cite{sen2}.
It still must be possible to see if  
$U(1)$ gauge field is no longer dynamical from string field theory.
To this end, it is of interest to calculate the two point functions of
all vector particles at the tachyon condensate using level
truncation. 
It is of interest to compare with the situation
in p-adic string theory \cite{ghosen2}.
 The kinetic terms for the
 translationaly zero modes of the soliton in p-adic string theory
 disappears after field redefinition at the new stationary
 point\footnote{The author thanks A. Sen for pointing this out}.

The organization of this paper is as follows. In section 2 we review
Witten's bosonic open string field theory \cite{wit} to set up our 
notations and
conventions. We use the formulation of describing string field theory
in arbitrary background field developed by \cite{pes} and \cite{sen3}.
Then we show the existence of the field
redefinition \eq{redef}.
We also identify the set of string field theory
transformations which correspond to the gauge transformation of the
$U(1)$ gauge field in the low energy effective action. 
These are shown to exist at the first non-linear order in
the fields. 
In section 3
we  review of how the 
the free action for the tachyon  and the gauge field
arises from string field theory.
In section 4 we obtain the effective action ${\cal S}(\ta_\mu,
\tilde{\phi})$ by eliminating all other massive and auxiliary fields.
In section 5 we derive
the constraint equations for field redefinition. 
In section 6 we  write down
the consistency conditions which should be
identically satisfied.
In section 7 we verify the consistency conditions
are identically satisfied and
in section 8 we show the existence of the solutions for the
consistency conditions and thus prove the existence of the field
redefinition up to the first nonlinear order in the fields.
This also identifies the string field theory transformation which
corresponds to the gauge transformation of the $U(1)$ field of the low
energy effective action.
In section 9 we examine the two point function of the transverse
photon. We show explictly that the kinetic term for the transverse
photon does not decrease as fast as the tachyon potential approaches
zero in the level truncation approximation. We show that at level
$(3,6)$ there is a physical vector excitation in the spectrum, but at
level $(4,8)$ this might be removed or pushed to a higher energy.

\section{Review of string field theory}

The open string field theory action is given by 
\be
{\cal S} = -\frac{1}{g^2} \left( \frac{1}{2}\langle I\circ \Phi(0)
Q_B \Phi(0) \rangle + \frac{1}{3} \langle f_1 \circ \Phi(0) f_1\circ
\Phi (0) f_3\circ\Phi (0) \rangle \right )
\ee
$g$ is the open string coupling constant.
$Q_B$ is the BRST charge, and by $\langle\,  .\,  \rangle$ we mean
correlation functions evaluated in the $SL(2,R)$ vacuum 
$\langle 0 |\,  . \, | 0\rangle$
Here $\Phi$ stands for the vertex operator in the conformal field
theory of the first quantized open string theory with ghost number
one. Using the standard state to operator mapping the field $\Phi$
creates the state $|\Phi\rangle$ out of the $SL(2,R)$ vacuum.
\be
|\Phi\rangle = \Phi(0)|0\rangle
\ee
As $\Phi$ has ghost number one, the Hilbert space of the first
quantized open string theory including the $b$ and $c$ ghost fields 
is restricted  to ghost number one. $f\circ \Phi(z)$ denotes the
conformal transformation of the field $\Phi$ by the map $f(z)$. For
example if $\Phi$ is a primary field of weight $h$ then $f\circ
\Phi(z) = (f^{\prime}(z))^h \Phi(f(z))$. The conformal transformations
$f_1, f_2,f_3$ and $I$ are are given below.
\bea
f_1(z) &=& -e^{-i\pi/3} \left. 
\left[ \left(\frac{1-iz}{1+iz}\right)^{2/3}
-1 \right]  \right/ 
\left[ \left(\frac{1-iz}{1+iz}\right)^{2/3} +
e^{i\pi/3} \right] \\ \nonumber
f_2(z) = F(f_1(z)),  &\;&  f_3(z) = F(f_2(z)), \\ \nonumber
F(z) = - \frac{1}{1+z}, &\;&
I(z) = -\frac{1}{z} 
\eea

It is convenient to work in a given basis of the Hilbert space of the
first quantized string theory with ghost number one. Therefore we
write
\be
|\Phi\rangle = \sum_{r} \psi_r|\Phi_{1;r}\rangle
\ee
$r$ denotes the various fields as well as momenta. The subscript $1$
in $|\Phi_{1;r}\rangle$ stands for the ghost number. In terms of this
basis the string field theory action is given by
\be
\label{action}
{\cal S} = -\frac{1}{g^2} \left( \frac{1}{2}
\taq_{rs}\psi_{r}\psi_{s} + \frac{1}{3} \tac_{rst}\psi_r\psi_s\psi_t
\right)
\ee
Here summation over the repeated indices is implied. $\taq_{rs}$ and
$\tac_{rst}$ are given by
\bea
\taq_{rs} &=& \langle I\circ \Phi_{1;r}(0) Q_B \Phi_{1,s}(0) \rangle
\\ \nonumber
\tac_{rst} &=& \langle f_1\circ\Phi_{1;r} (0) f_2\circ\Phi_{1,s} (0)
f_3 \circ\Phi_{1,t} (0)\rangle
\eea
The vertices $\taq_{rs}$ and $\tac_{rst}$ have the following symmetry
property
\bea
\taq_{rs} = \taq_{sr} \\ \nonumber
\tac_{rst}= \tac_{str} =\tac_{trs}
\eea

The string field theory action has a large gauge symmetry. The
infinitesimal gauge transformation law of $\psi_s$ is given by
\be
\label{gauge}
\delta\psi_r = \aq_{r\alpha}\lambda_{\alpha} + \ac_{r\alpha
s}\lambda_{\alpha}\psi_s
\ee
where
\bea
\label{degauge}
\aq_{r\alpha} &=& \langle \Phi^{c}_{2;r}|Q_B | \Phi_{0;\alpha}\rangle
\\ \nonumber
\ac_{r\alpha s} &=& \left( 
\langle f_1 \circ \Phi^c_{2;r}(0) f_2\circ
\Phi_{1;s}(0) f_3\circ\Phi_{0; \alpha}(0) \rangle \right. \\ \nonumber
&-& 
\left. \langle f_1 \circ \Phi^c_{2;r}(0) f_2\circ
\Phi_{0;\alpha}(0) f_3\circ\Phi_{1;s}(0)  \rangle \right) (-1)^{n_r +
n_r^g}
\eea
where $n_r$ and $n_r^g$ denotes the fact that $|\Phi_{1;r}\rangle$ 
is an $n_r$th level secondary in the matter sector, and $n_r^g$th
level secondary in the ghost sector. $\langle \Phi_{2;r}^c|$ is
conjugate to the state $|\Phi_{1;r}\rangle$, that is
\be
\langle \Phi_{2;r}^c| \Phi_{1;s} \rangle = \delta_{rs}
\ee
In addition to  the gauge symmetries, the string field theory
action has a `trivial' symmetry of the form
\be
\delta\psi_r = K_{rs}(\{ \psi \})\frac{ \delta {\cal S} }{\delta\psi_s}
\ee
where $K_{rs}$ is an antisymmetric in $r$ and $s$ and is a function of
the components $\psi_r$.
It is easy to verify that this is a symmetry of the action. This
symmetry will play an important role in the field redefinition.

\section{Free action of the tachyon and the gauge field}

In this section we review how to obtain
the free action of the tachyon and the gauge
field as a warm up exercise for the next sections.
We work out the action for the first two levels in the expansion of
$\Phi$.
\be
|\Phi\rangle = \int dk 
\tphi(k) c_1|k\rangle + \int dk \ta_\mu
(k) c_1 \alpha_{-1}^\mu (0)  |k\rangle + 
\int dk F(k)c_0  |k\rangle
\ee
Here $\tphi$ corresponds to the tachyon, $\ta_\mu$ the $U(1)$ gauge
field. We will see below that $F(k)$ an auxiliary field. 
By $\int dk$ we mean the $\int d^{26}k$, the state $|k\rangle$ denotes
$e^{ikX(0)}|0\rangle$.
Substituting this expansion in
\eq{action} and keeping only the quadratic vertices we obtain the
following action for the first two levels.
\bea
\label{free}
{\cal{S}}
= -\frac{1}{g^2} \left[ \int dk  \left( \tphi(-k)(2k^2 -1) \tphi(k) + 
\ta_\mu(-k)
(2k^2) \ta^\mu(k) \right.\right.  \\ \nonumber
\left.\left. - 2F(-k)F(k) +4k^\mu\ta_\mu (-k) F(k) \right)
 \right]
\eea
We are working in the units $\alpha' =2$.
The gauge transformation of these fields can be determined by using
\eq{gauge}. Restricting to
only the linearized gauge transformation that is, 
only terms from $\aq_{r\alpha}$ in \eq{gauge} we obtain
\bea
\label{freegauge}
\delta \tphi(k) &=& 0 \\ \nonumber
\delta \ta_\mu (k) &=& -2k_\mu \lambda(k) \\ \nonumber
\delta F(k) &=& 2k^2 \lambda(k)
\eea
Here $\lambda(k)$ is the infinitesimal gauge parameter. The quadratic
piece of the action in \eq{action} is invariant under linearized gauge
transformation. It is easy to verify the action in \eq{free} is
invariant under the transformation of \eq{freegauge}

The auxiliary field $F(k)$ can be eliminated using the equations of
motion
\be
F(k) = -k^\mu \ta_\mu(k)
\ee
On eliminating $F(k)$, the action reduces to that of the tachyon with
the standard kinetic term for the $U(1)$ gauge field. It is given by
\be
{\cal S } =
-\frac{1}{g^2}
\left[ \int dk   \tphi(-k)(2k^2 -1) \tphi(k) + \ta_\mu(-k)
(2k^2) \ta^\mu(k) - 2 \ta_\mu(-k) k^\mu k^\nu \ta_\nu(k) \right]
\ee
Let us now  define the gauge parameter $i\eta(k) = 2\lambda(k)$
, the gauge transformation of
the $U(1)$  field $\ta_\mu$(k) is given by
\be
\delta\ta_\mu(k) = ik_\mu \eta(k)
\ee
Now it is clear that at the free level the tachyon $\tilde{\phi}$ and
the gauge field $\ta_\mu (k)$ can be identified with the tachyon
$\phi(k)$ and the gauge field $A_\mu(k)$ of the low energy effective
action.

\section{Elimination of the massive and auxiliary fields}

Let us first obtain the low energy effective action  in terms of
the tachyon $\tphi$ and the $U(1)$ gauge field $\ta_\mu$. We do this
by eliminating all other
the massive fields and auxiliary fields in the
string field $\Phi$ using their equations of motion. Of course such
an effective action is valid only to tree level. 

We divide the string fields $\psi_{r}$ into two kinds.
$\psi_{r}, \psi_{s} \ldots $ stand for the tachyon $\tphi(k)$ and 
$\ta_\mu(k)$ including their momentum index. The fields $\psi_{a},
\psi_{b} \ldots$ stand for higher string modes, auxiliary fields
and their momenta. Solving for $\psi_a$ in terms of $\psi_r$ up to the
first non-linear order in the fields we obtain
\be
\psi_a = \el_{ar}\psi_r + \frac{1}{2}\es_{ars}\psi_r\psi_s
+ O (\psi^3)
\ee
where $\el_{ar}$ and $\es_{ars}$ are determined from the equation of
motion for $\psi_a$. $\es_{ars}$ is symmetric in $r$ and $s$, by
definition.
They are given by solving the following equations
\bea
\label{eqom}
\taq_{ar} + \taq_{ab}\el_{br} &=&0 \\ \nonumber
\frac{1}{2}\taq_{ab}\es_{brs} + 
\tac_{ars} + \tac_{abs}\el_{br} &+& \tac_{arb}\el_{bs} +
\tac_{abc}\el_{br}\el_{cs} = 0
\eea
To solve these equations we need $\taq_{ab}$ to be invertible. The zero
eigen modes of this operator arises from BRST exact states. So we
restrict ourselves to states which are not BRST exact to ensure that
the operator $\taq_{ab}$ is invertible.
The linearized equation of motion for $\psi_r$ now becomes
\be
\taq_{sr}\psi_{r} + \taq_{sa}\el_{ar}\psi_{r} = 0
\ee
From the gauge transformation of $\psi_a$ at the linear order we
obtain the relation
\be
\label{gaugerelation}
\aq_{a\alpha} = \el_{ar}\aq_{r\alpha}
\ee
The gauge transformation of $\psi_r$ up to the first nonlinear order
can be written down as
\be
\label{origauge}
\delta \psi_r = \aq_{r\alpha}\lambda_\alpha + (\ac_{r\alpha s} +
\ac_{r \alpha a} \el_{a s} ) \lambda_\alpha\psi_s + O (\psi^2)
\ee
It is clear from this gauge transformation, that $\psi_r$ does not
have the same 
gauge transformation property of the $U(1)$ gauge field or
that of the tachyon, because of the non-linear terms. The 
conventional low
energy effective action of open bosonic string theory has the
tachyon denoted by $\phi(k)$
and the gauge field, $A_\mu(k)$, 
The action should be invariant under the
following transformation
\be
\label{gaugephi}
\delta \phi (k) =0 \;\;\;  \delta A_\mu = ik_\mu \eta(k),
\ee
where $\eta(k)$ is the gauge transformation parameter. Let us call
these low energy fields as $\phi_i$. The index $i$ labels  the
tachyon, the gauge field and their momenta. 
Let the gauge parameters of these fields be labelled by
$\eta_{\kappa}$.
It must be possible to
relate $\psi_r$ to $\phi_i$ so that $\phi_i$  has the gauge
transformation given by \eq{gaugephi}. It also must be possible to
relate the $U(1)$ gauge invariance of the low energy action to a a
string field theory gauge transformation.
We show that this is possible
in the next section. We use the method developed by \cite{ghosen} for
field redefinition in the closed non-polynomial string field theory. We
account for the elimination of all the massive and auxiliary fields
and the fact that we are dealing with open string field theory. We
verify the consistency conditions for field redefinitions in open
string field theory.

\section{Field redefinition}

In this section we find the conditions which need to be satisfied to
redefine the fields that appear in the string field 
theory action to that
which appear in the low energy effective action.
Let us relate $\psi_r$ and $\phi_i$ by the following general formula
up to the first non linear order
\be
\label{define}
\psi_r = \cl_{ri}\phi_i + \frac{1}{2}\cq_{rij} \phi_i \phi_j
+ O(\phi^3)
\ee
The gauge parameter $\lambda_\alpha$ is related to the 
gauge parameters of  the low energy fields by the following 
general formula
\be
\label{defgauge}
\lambda_\alpha = \dl_{\alpha\kappa}\eta_{\kappa} + \dq_{\alpha\kappa
i} \eta_{\kappa} \phi_i + O(\phi)
\ee
In addition to the gauge symmetry of the string field theory action,
the action is also invariant under the transformation
\bea
\label{extra}
\delta\psi_{r}&=& K_{rs}\frac{\delta {\cal S} }{\delta \psi_s} +
K_{ra}\frac{\delta {\cal S} }{\delta \psi_a} \\ \nonumber
 &=& K_{rs} (\taq_{st}\psi_t + \taq_{sa} \el_{at} \psi_t )
 + O(\psi^2)
\eea
Where we have used the fact that we have eliminated the fields $\psi_a$
using the equation of motion. We have  kept terms only up to linear
order in the fields. We now assume the most general form for the
function $K_{rs}$
\be
\label{extrad}
K_{rs} = \kl_{rs\kappa}\eta_\kappa + O(\phi)
\ee
The low energy fields $\phi_i$ have the following gauge transformation
property
\be
\label{lowgauge}
\delta\phi_i = \bl_{i\kappa}\eta_\kappa
\ee
Substituting \eq{lowgauge} in the variation obtained from \eq{define}
we get
\be
\label{var1}
\delta\psi_r = \cl_{ri}\bl_{i\kappa} \eta_{\kappa} +
\cq_{rij}\phi_i\bl_{j\kappa}\eta_{\kappa}
\ee
Now from \eq{origauge}, \eq{gauge} , \eq{extra}, \eq{extrad} we obtain
\bea
\label{var2}
\delta\psi_r&=& \aq_{r\alpha}\dl_{\alpha\kappa}\eta_\kappa +
\aq_{r\alpha}\dq_{\alpha\kappa i}\eta_{\kappa}\phi_i + (\ac_{r\alpha
s} + \ac_{r\alpha
a}\el_{as})\dl_{\alpha\kappa}\cl_{si}\eta_\kappa\phi_i  \\ \nonumber
&+&
\kl_{rs\kappa} (\taq_{st} + \taq_{sa}\el_{at})
\cl_{ti}\eta_{\kappa}\phi_i
\eea
From comparing \eq{var1} and \eq{var2} we get the following set of
equations
\bea
\label{consist1}
\cl_{ri}\bl_{i\kappa}\et{\kappa}{\rho} &=& 
\aq_{r\alpha}\dl_{\alpha\kappa}\et{\kappa}{\rho}
\\ 
\label{consist2}
\cq_{rij}\ph{i}{m}\bl_{j\kappa}\et{\kappa}{\rho} &=&
\aq_{r\alpha}\dq_{\alpha\kappa i}\et{\kappa}{\rho}\ph{i}{m} + 
(\ac_{r\alpha
s} + \ac_{r\alpha
a}\el_{as})\dl_{\alpha\kappa}\cl_{si}\et{\kappa}{\rho}
\ph{i}{m}  \\ \nonumber
&+&
\kl_{rs\kappa} (\taq_{st} + \taq_{sa}\el_{at})
\cl_{ti}\et{\kappa}{\rho}\ph{i}{m}
\eea
Where we have introduced a complete set of gauge transformations 
$\{ \et{\kappa}{\rho} \}$ and a complete set of field configurations
$\{ \ph{i}{m} \}$
We have already seen in the previous section 
that we are able to find $\cl_{ri}$ and
$\dl_{\alpha\kappa}$ such that \eq{consist1} is satisfied.
It remains to be seen that if one can find $\cq_{rij}$, 
$\dq_{\alpha\kappa i}$  and $\kl_{rs\kappa}$
such that \eq{consist2} can be satisfied. 
It is easy to extend these constraint equations to higher orders in
fields.
There
are obstructions to this. 
These obstructions can arise if under some conditions \eq{consist2} 
reduces to equations involving known quantities which should be 
satisfied identically. 
In the next section we will enumerate them.

\section{Consistency conditions for field redefinition}

There are three consistency conditions  in all for the 
\eq{consist2}. We will
enumerate each of them in the following subsections. These are
similar to the consistency conditions 
found in \cite{ghosen} for the
closed string. The difference being that we have eliminated all the
auxiliary and massive fields using equations of motion and we are
dealing with open string field theory.

\subsection{Condition A}
Divide the complete set of fields $\{\ph{i}{m}\}$ to $2$ sets
$\{\hph{i}{\hat{m}}\}$ and $\{\bph{i}{\bar{m}}\}$. The set 
$\{\hph{i}{\hat{m}}\}$
satisfies linearized equation of motion and the 
set $\{\bph{i}{\bar{m}}\}$ 
does not. 
\bea
\label{space1}
(\taq_{st} + \taq_{sa}\el_{at} ) \cl_{ti}\hph{i}{\hat{m}} &=& 0 
\;\;\; \hbox{for all} \; s \\ \nonumber
(\taq_{st} + \taq_{sa}\el_{at} ) \cl_{ti}\bph{i}{\bar{m}} &\neq& 0
\;\;\; \hbox{for some} \; s 
\eea
Now divide the complete set of gauge transformations
$\{\et{\kappa}{\rho}\}$ to sets $\{\het{\kappa}{\hat{\rho}}\}$ and
$\{\bet{\kappa}{\bar{\rho}}\}$ such that
\bea
\label{space2}
\bl_{i\kappa}\het{\kappa}{\hat{\rho}} &=& 0 \;\;\; 
\hbox{for all} \; i \\ \nonumber
\bl_{i\kappa}\bet{\kappa}{\bar{\rho}} &\neq& 0 \;\;\; 
\hbox{for some} \; i
\eea
Thus $\{\het{\kappa}{\hat{\rho}}\}$ are those gauge transformations for 
which there is no variation in the fields $\phi_i$.
The index $r$ is divided also to two sets into BRST invariant and
non-invariant states given by
\be
\label{space3}
\langle \hat{\Phi}^c_{2;r} | Q_B =0 \;\;\; \langle
\bar{\Phi}^c_{2;\bar{r} }| Q_B \neq 0
\ee

Now restrict the indices $r, m, \rho$ in \eq{consist2} to $\hat{r},
\hat{m}$ and $\hat{\rho}$. The terms involving $\cq$ and $\kl$ drop out
because of \eq{space1} and \eq{space2} respectively. The term
involving $\dq$ drops off because of
\be
\label{space4}
\aq_{\hat{r}\alpha} =0
\ee
This follows from the definition of $\aq$ in \eq{gauge} and
\eq{space3}. Thus we obtain the consistency condition
\be
\label{obs1}
(\ac_{r\alpha s} + \ac_{r\alpha a} \el_{a s} ) \dl_{\alpha \kappa}
\cl_{si} \het{\kappa}{\hat{\rho}}\hph{i}{\hat{m}} = 0
\ee
This involves all known coefficients and should be satisfied
identically. 

\subsection{Condition B}
The next obstruction arises because of  $\cq_{rij}$ is symmetric in
$i$ and $j$. Restrict the index $r$ in \eq{consist2} to be $\hat{r}$.
The term involving $\dq$ drops out because of \eq{space4}. Take
$\ph{j}{m}$ to be of the form
$\bl_{j\kappa^{\prime}}\et{\kappa^{\prime}}{\rho^{\prime}}$. Then term
involving $\kl$ is given by
\bea
\kl_{\hat{r}s\kappa}
(\taq_{st} &+& \taq_{sa}\el_{at})\cl_{ti}\et{\kappa}{\rho}
\bl_{j\kappa^{\prime}}\et{\kappa'}{\rho'} 
=  \\ \nonumber
&=& \kl_{\hat{r}s\kappa}(\taq_{st} +
\taq_{sa}\el_{at})
\et{\kappa}{\rho}\aq_{t\alpha}
\dl_{\alpha\kappa^{\prime}} \et{\kappa'}{\rho'} \\ \nonumber
&=&
\kl_{\hat{r}s\kappa}
(\taq_{st}\aq_{t\alpha} + \taq_{sa}{\aq_{a\alpha}} )
\et{\kappa}{\rho}\aq_{t\alpha}
\dl_{\alpha\kappa^{\prime}} \et{\kappa'}{\rho'} \\ \nonumber
&=&0
\eea
Where we have used \eq{consist1} in the first step and
\eq{gaugerelation} in the second step. We have
\be
\label{gaugeinv}
\taq_{st}\aq_{t\alpha} + \taq_{sa}{\aq_{a\alpha}} = \langle 
I\circ\Phi_{1;s} Q_B^2 \Phi_{0;\alpha}\rangle =0
\ee
To arrive at the above relation we have used completeness and
$Q_B^2 =0$. We also use the 
fact correlations vanish except when saturated by
fields with total ghost number three. 
The term involving $\kl$ also vanishes. 
Thus for this case from \eq{consist2} we are left with 
\be
\cq_{\hat{r}ij}\bl_{i\kappa'}\et{\kappa'}{\rho'}
\bl_{j\kappa}\et{\kappa}{\rho} = 
(\ac_{\hat{r} \alpha s} + \ac_{\hat{r} \alpha a}\el_{a s} )
\dl_{\alpha\kappa}\cl_{si}
\bl_{i\kappa'} \et{\kappa}{\rho}
\et{\kappa'}{\rho'}
\ee
As $\cq_{\hat{r}ij}$ is symmetric in $i$ and $j$ we have the following
constraint
\be
\label{obs2}
(\ac_{\hat{r} \alpha s} + \ac_{\hat{r} \alpha a}\el_{a s} )
\dl_{\alpha\kappa}\cl_{si}
\bl_{i\kappa'} (
\et{\kappa}{\rho} \et{\kappa'}{\rho'}-
\et{\kappa'}{\rho'} \et{\kappa}{\rho}) =0
\ee

\subsection{Condition C}
We now find the third and final obstruction. This is due to the
antisymmetry of $\kl_{rs\alpha}$ in $r$ and $s$. In \eq{consist1}
choose the index $\rho$ to be $\hat{\rho}$. Then by \eq{space2} the
term involving $\cq$ in \eq{consist1} drops out. Now multiply the
equation by $(\taq_{rt'} + \el_{ar}\taq_{at'})\cl_{t'j}\ph{j'}{m'}$
The term involving $\dq$ is then given by
\bea
\aq_{r\alpha}(\taq_{rt'} + \el_{ar}\taq_{at'})
\cl_{t'j'}\ph{j'}{m'}\dq_{\alpha\kappa i}
\het{\kappa}{\hat{\rho}}\ph{i}{m} \\ \nonumber
= (\aq_{r\alpha}\taq_{rt'} + \aq_{a\alpha}\taq_{at'} ) 
\cl_{t'j'}\ph{j'}{m'}\dq_{\alpha\kappa i}
\het{\kappa}{\hat{\rho}}\ph{i}{m} 
\eea
Where we have used \eq{gaugerelation}. Now using the fact that $\taq$
is a symmetric matrix and \eq{gaugeinv}, the term involving $\dq$
vanishes. Thus \eq{consist2} becomes 
\bea
(\ac_{r\alpha s} + \ac_{r\alpha a}\el_{as})
(\taq_{rt'} + \el_{ar}\taq_{at'})
\cl_{t'j'}\dl_{\alpha\kappa}\cl_{si}\het{\kappa}{\hat{\rho}}
\ph{j'}{m'}\ph{i}{m} 
+ \\ \nonumber 
\kl_{rs\kappa}
(\taq_{st} + \taq_{sa}\el_{at})\cl_{ti}
\het{\kappa}{\hat{\rho}}\ph{i}{m}
(\taq_{rt'} + \taq_{rb}\el_{bt'})\cl_{t'j}\het{\kappa'}{\hat{\rho}'}
\ph{i'}{m'} =0
\eea
As $\kl_{rs\alpha}$ is antisymmetric in $r$ and $s$, 
the symmetric component in $r$ and $s$ should be zero.
Therefore we  have the
following constraint
\be
\label{obs3}
(\ac_{r\alpha s} + \ac_{r\alpha a}\el_{as})
(\taq_{rt'} + \el_{ar}\taq_{at'})
\cl_{t'j'}\dl_{\alpha\kappa}\cl_{si}\het{\kappa}{\hat{\rho}}
(\ph{j'}{m'}\ph{i}{m} + \ph{j'}{m}\ph{i}{m'}) =0
\ee

\section{Verification of the consistency conditions}

In this section we show that each of the consistency conditions 
found in the
previous section is satisfied identically for the open superstring
theory.

\subsection{Condition A}
Let us first verify that equation \eq{obs1} is satisfied.
Define
\be
\label{d1}
|\hat{\Psi}^{\hat{(m)}}_{1} \rangle =
\cl_{si}\hph{i}{\hat{m}}|\Phi_{1;s}\rangle +
\el_{as}\cl_{si}\hph{i}{\hat{m}}|\Phi_{1;a}\rangle
\ee
Now
\be
\langle\Phi_{1;r}|Q_B|\hat{\Psi}^{\hat{(m)}}\rangle = 
\cl_{si}\hph{i}{\hat{m}}(\taq_{rs} +\taq_{ra}\el_{as}) =0
\ee
Where we have used \eq{space1}. Also from \eq{eqom} we have
\be
\langle\Phi_{1;a}|Q_B|\hat{\Psi}^{\hat{(m)}}\rangle = 
\cl_{si}\hph{i}{\hat{m}}(\taq_{as} +\taq_{ab}\el_{bs}) =0
\ee
This implies that $|\hat{\Psi}^{\hat(m)} \rangle$ is a BRST invariant
state.
\be
Q_B|\hat{\Psi}^{\hat{(m)}}\rangle =0
\ee
Now define 
\be
\label{d2}
|\hat{\Lambda}^{\hat{(\rho)}}\rangle =
\dl_{\alpha\kappa}\het{\kappa}{\hat{\rho}} |\Phi_{0;\alpha}\rangle
\ee
We use \eq{consist1} and \eq{space2} to show
\bea
\label{brst1}
\langle\Phi^{c}_{2;r} | Q_B | \hat{\Lambda}^{\hat{(\rho)}}\rangle
&=& \dl_{\alpha\kappa}\het{\kappa}{\hat{\rho}}
\langle\Phi_{2;r}^c|Q_B|\Phi_{0\alpha}\rangle  \\ \nonumber
&=&
\dl_{\alpha\kappa}\het{\kappa}{\hat{\rho}}\aq_{r\alpha} \\ \nonumber
&=& \cl_{ri}\bl_{i\kappa}\het{\kappa}{\hat{\rho}} \\ \nonumber 
&=&0
\eea
Also we have 
\bea
\label{brst2}
\langle\Phi^{c}_{2;a} | Q_B | \hat{\Lambda}^{\hat{(\rho)}}\rangle
&=& \dl_{\alpha\kappa}\het{\kappa}{\hat{\rho}}
\langle\Phi_{2;a}^c|Q_B|\Phi_{0\alpha}\rangle  \\ \nonumber
&=&
\dl_{\alpha\kappa}\het{\kappa}{\hat{\rho}}\aq_{a\alpha} \\ \nonumber
&=& \cl_{ri}\bl_{i\kappa}\het{\kappa}{\hat{\rho}}\el_{ra} \\ \nonumber 
&=&0
\eea
Where we have used \eq{gaugerelation} and \eq{space2}.
From \eq{brst1} and \eq{brst2}  we obtain that
$|\hat{\Lambda}^{\hat{(\rho)}}\rangle$ is a BRST invariant state.
\be
Q_B|\hat{\Lambda}^{\hat{(\rho)}}\rangle =0
\ee
Using definitions \eq{d1}, \eq{d2} and \eq{degauge} the  
consistency condition \eq{obs1} reduces to
\be
\label{consist4}
\langle f_1\circ\hat{\Phi}_{\hat{r}}^{c}
f_2\circ\hat{\Lambda}^{\hat{(\rho)}} 
f_3\circ \hat{\Psi}^{\hat{(m)}}
\rangle -
\langle f_1\circ\hat{\Phi}_{\hat{r}}^{c}
f_2\circ \hat{\Psi}^{\hat{(m)}}
f_3\circ\hat{\Lambda}^{\hat{(\rho)}} 
\rangle =0
\ee
As all the field in \eq{consist4} are BRST invariant fields, if any of
them is BRST exact, then the equation is automatically satisfied.
Therefore we have to look for BRST invariant but not BRST exact
fields.
The linearized gauge transformation parameter 
for the tachyon field is zero. Thus by standard BRST analysis the only
gauge transformation we have is that for the photon. This is given by
$
|\hat{\Lambda}^{\hat{(\rho)}} \rangle = \xi|0\rangle
$
The ghost number one fields in the formula \eq{consist4} can by the on
shell tachyon 
$ |\hat{\Psi}^{\hat{(m)}}\rangle = c_1|k\rangle $
with $k^2=1/2$, the mass shell condition. Or it is the on shell gauge
field given by
$|\hat{\Psi}^{\hat{(m)}}\rangle = \epsilon_{\mu}c_1
\alpha_{-1}^{\mu} |k\rangle
$
with $k^\mu\epsilon_\mu =0$ and $k^2=0$.  The conjugate fields in
\eq{consist4} are the fields conjugate to the on shell tachyon and the
on shell gauge field. We can have
$
\langle \Phi_{\hat{r}}^{c}|  =  
\langle k|c_{-1}c_0 
$
with $k^2=1/2$, or we have
$
\langle \Phi_{\hat{r}}^{c} | =
\langle k| c_{-1}c_0\alpha_{-1}^{\mu}a_\mu
$
with $a_\mu k^\mu=0$ and $k^2=0$. Evaluating the left hand side of
\eq{consist4} with any of these states it can be seen that it is
zero. Thus the consistency condition 
\eq{obs1} is identically satisfied.

\subsection{Condition B}

Now let us examine the obstruction\eq{obs2}. Define
$|\Lambda^{(\rho)}\rangle 
= \dl_{\alpha\kappa}\et{\kappa}{\rho}|\Phi_{0;\alpha}\rangle$. 
Then we
have
\bea
\cl_{si}\bl_{i\kappa'}\et{\kappa'}{\rho'}|\Phi_{1;s}\rangle &+&
\el_{as}\cl_{si}\bl_{i\kappa'}\et{\rho'}{\kappa'}|\Phi_{1;a}\rangle
 \\ \nonumber
&= &
\aq_{s\alpha}\dl_{\alpha\kappa'}\et{\kappa'}{\rho'}|\Phi_{1;s}\rangle
+
\el_{as}\aq_{s\alpha}\dl_{\alpha\kappa'}\et{\kappa'}{\rho'}
|\Phi_{1;a}\rangle \\ \nonumber
&=& \langle\Phi_{2;s}^{c}|Q_B|\Phi_{0;\alpha}\rangle
\dl_{\alpha\kappa'} \et{\kappa'}{\rho'}|\Phi_{1;s}\rangle +
\langle
\Phi_{2;a}^c|Q_B|\Phi_{0\alpha}\rangle\dl_{\alpha\kappa'}
\et{\kappa'}{\rho'} | \Phi_{1;a}\rangle \\ \nonumber
&=& \dl_{\alpha\kappa'}\et{\kappa'}{\rho'}Q_B|\Phi_{0;\alpha}\rangle
\\ \nonumber
&=& Q_B|\Lambda^{(\rho')}\rangle
\eea
Here we have used \eq{consist1}, \eq{degauge} and completeness.  
Thus the consistency condition \eq{obs2} can be written as
\be
\langle f_1 \circ\Phi_{\hat{r}}^c (0)
f_2\circ \Lambda^{(\rho)}(0) f_3\circ Q_B\Lambda^{(\rho')}(0) \rangle
- \langle f_1 \circ\Phi_{\hat{r}}^c (0)
f_2\circ Q_B \Lambda^{(\rho')}(0) f_3 \Lambda^{(\rho)}(0) \rangle
- (\rho \leftrightarrow \rho')
\ee
Now, one can deform the contour of $Q_B$ in the first set of term so
that it will act on $\Lambda^\rho$. This is because
$|\Phi_{\hat{r}}^c\rangle$ is BRST invariant. 
Then, the first set of term
cancel against the second set of terms. Thus the consistency condition
\eq{obs2} is satisfied identically.

\subsection{Condition C}
Finally we analyze the third consistency condition \eq{obs3}.
Define 
\be
|\Psi^{(m)}\rangle = \cl_{tj}\ph{j}{m}|\Phi_{1,t}\rangle +
\el_{at}\cl_{tj}\ph{j}{m}|\Phi_{1,a}\rangle  =
\psi_{t}^{(m)}|\Psi_{1;t}\rangle + 
\psi_{a}^{(m)}|\Psi_{1;a}\rangle 
\ee
and 
$ |\Lambda^{\rho}\rangle = \dl_{\alpha\kappa}\et{\kappa}{\rho}|
\Phi_{0;\alpha}\rangle$. Using these definitions $\eq{obs3}$ can be
written as 
\bea
\label{add1}
\ac_{r\rho s}\taq_{rt'} \psi_{t'}^{(m')} \psi_{s}^{(m)} &+&
\ac_{r\rho a}\taq_{rt'} \psi_{t'}^{(m')} \psi_{a}^{(m)} + \\
\nonumber
\ac_{r\rho s}\taq_{ra'} \psi_{a'}^{(m')} \psi_s^{(m)} &+&
\ac_{r\rho a}\taq_{ra'} \psi_{a'}^{(m')} \psi_a^{(m)} +
(m\leftrightarrow m') =0
\eea
Where we have used the linearized 
equation of motion of $\psi_a^{(m)}$ and \eq{eqom}. 
Now one can add to this equation the following equation 
\bea
\label{add2}
\ac_{b\rho s}\taq_{bt'} \psi_{t'}^{(m')} \psi_{s}^{(m)} &+&
\ac_{b\rho a}\taq_{bt'} \psi_{t'}^{(m')} \psi_{a}^{(m)} + \\
\nonumber
\ac_{b\rho s}\taq_{ba'} \psi_{a'}^{(m')} \psi_{s}^{(m)} &+&
\ac_{b\rho a}\taq_{ba'} \psi_{a'}^{(m')} \psi_{a}^{(m)} +
(m\leftrightarrow m') =0
\eea
This equation holds because $\psi_{a}^{(m)}$ satisfies linearized
equation of motion.
\be
\taq_{ar}\psi_{r}^{(m)} + \taq_{ab}\psi_{b}^{(m)} =0
\ee
We are justified in using the linear equations of motion because the
terms in \eq{obs3} are quadratic in the fields. Adding \eq{add1} and
\eq{add2} and using completeness, the constraint \eq{obs3} becomes
\bea
\left( \langle f_1\circ Q_B \Psi^{(m)}(0)  f_2\circ
\hat{\Lambda}^{\hat{(\rho)}} (0) f_3\circ \Psi^{(m')}(0)\rangle \right.
&-& \\ \nonumber
\left. \langle f_1\circ Q_B \Psi^{(m)}(0)  
f_2\circ \Psi^{(m')} (0) f_3\circ \hat{\Lambda}^{\hat{(\rho)}} (0) 
\rangle  \right) 
+ (m\leftrightarrow m') =0
\eea

One can now deform the contour of $Q_B$ in the first set of terms 
so that it acts on $|\Psi^{(m')}\rangle$. It does not act on on
$|\hat{\Lambda}^{\hat{(\rho)}}\rangle$ as it is BRST invariant. In
deforming the contour one picks up minus sign, and another minus sign
as $Q_B$ crosses $\Psi^{(m)}$. Now these first set of
terms cancel against the second set of terms. Thus \eq{obs3} is
satisfied identically.

We have shown that all the consistency conditions 
found in the previous section
for solving \eq{consist2} are satisfied identically.

\section{Existence of field redefinition} 

In this section we show that there exists $\cq$ and $\dq$ which solve
\eq{consist2}. This establishes that there exists a field redefinition
such that in the new set of fields the tachyon and the $U(1)$ gauge
field corresponds to the fields seen in the low energy effective
action to the first non-linear order in the fields. This also
establishes the existence of a string gauge transformations which
corresponds to the $U(1)$ gauge transformation of the low energy
effective action. 

In this  section we will closely follow \cite{ghosen}. 
We outline the steps involved
for completeness. We divide the situations to three cases

\subsection{Case A: $r\in \bar{r}$}

Let us consider the case when $r$ belongs to the type $\bar{r}$. 
This
implies that $\aq_{\bar{r} \alpha}$ has a right inverse $M_{\alpha
\bar{s}}$ 
\be
\aq_{\bar{r}\alpha}M_{\alpha\bar{s}} = \delta_{\bar{r}\bar{s}}
\ee
Therefore we can solve \eq{consist2} for this case by choosing
$\dq$ given by
\bea
\dq_{\alpha\kappa i}\eta_{\kappa}\phi_i &=&
M_{\alpha\bar{r}}\left(
\cq_{\bar{r}ij}\phi_i\bl_{j\kappa}\eta_\kappa -
\aq_{\bar{r}\beta}\dq_{\beta\kappa i}\eta_{\kappa}\phi_i  \right. \\
\nonumber
&-& \left. (\ac_{\bar{r}\beta
s} + \ac_{r\beta
a}\el_{as})\dl_{\beta\kappa}\cl_{si}\eta_\kappa\phi_i 
+
\kl_{\bar{r}s\kappa} (\taq_{st} + \taq_{sa}\el_{at})
\cl_{ti}\et{\kappa}{\rho}\ph{i}{m}  \right)
\eea

\subsection{Case B: $r\in \hat{r}$, $\rho\in \bar{r}$}

In this case the term involving $\dq$ in \eq{consist2} vanishes
because of \eq{space3}. Since the index 
$\rho$ is in the set $\bar{\rho}$ the matrix $S_{i\bar{\rho}} =
\bl_{i\kappa}\bet{\kappa}{\bar{\rho}}$ has a left inverse
$N_{\bar{\rho} j}$  which satisfies
\be
N_{\bar{\rho}i}S_{i\bar{\rho}'} = \delta_{\rho\bar{\rho}'}
\ee
We can now solve \eq{consist2} by choosing $\cq$ given by
\be
\cq_{\hat{r}ij} = \left(
 (\ac_{\hat{r}\alpha
s} + \ac_{\hat\alpha
a}\el_{as})\dl_{\alpha\kappa}\cl_{sj} +
\kl_{\hat{r}s\kappa} (\taq_{st} + \taq_{sa}\el_{at})
\cl_{tj}\right) \bet{\kappa}{\bar{\rho}} N_{\bar{\rho}i}
\ee
It  can be shown \cite{ghosen} that one can choose 
a basis for the fields $\{\phi_i\}$ such that the $\cq$ obtained from
the above equation is symmetric in $i$ and $j$.

\subsection{Case C: $r\in \hat{r}$, $\rho\in \hat{\rho}$, $m \in
\bar{m}$ }

The case $r\in \hat{r}$, $\rho\in \hat{\rho}$ and $m \in \hat{m}$ is
already covered in \eq{obs1}. We have shown in such a case the
consistency condition is satisfied identically. So, the case which
remains is $r\in \hat{r}$, $\rho\in \hat{\rho}$ and $m \in\bar{m}$. 
In this case the terms involving $\cq$ and $\dq$ drop from
\eq{consist2} due to \eq{space2} and \eq{space3} respectively. The
matrix $T_{s\bar{m}}$ given by
\be
T_{s\bar{m}} =
(\taq_{st} +
\taq_{sa}\el_{at})\cl_{tj}\bph{i}{\bar{m}}
\ee
Then the matrix $T_{s\bar{m}}$  has a left inverse $U_{\bar{m} r}$
given by
\be
U_{\bar{m} r} T_{r\bar{n}} = \delta_{\bar{m}\bar{n}}
\ee
Then we can solve \eq{consist2} by choosing 
\be
\kl_{\hat{r} t \kappa}\het{\kappa}{\hat{\rho}} =  - 
(\ac_{\hat{r}\alpha s} +\ac_{\hat\alpha
a}\el_{as})\dl_{\alpha\kappa}\cl_{sj}\het{\kappa}{\hat{\rho}}
\bph{i}{\bar{m}} U_{\bar{m} t}   
\ee
Notice that only the  projection of $\kl_{\hat{r} t \kappa}$ onto the
space spanned by $\het{\kappa}{\hat\rho}$ is determined. The other
components can be chosen to be arbitrary. It can also be shown that the
$\kl_{rs\kappa}$ can be chosen in such a way  that it is
antisymmetric in $r$ and $s$ \cite{ghosen}.

This completes the proof of the existence of field redefinition of the
tachyon and the gauge field so that they can be identified with the
fields that appear in the low energy effective action. This also
determines a string gauge transformation, $\lambda_{\alpha}$ which
corresponds to the $U(1)$ gauge transformation, $\eta_{\kappa}$ of the
low energy fields.

\section{The transverse photon at the tachyon condensate}

In this section we
demonstrate explictly that field redefinition is important in the
relation of the Dirac-Born-Infeld action and the low energy effective
action derived from Witten's String field theory. To do this we
evaluate the kinetic term for the transverse photon $A_\mu^T$,
to level 4 in the tachyon condensate. We work in the Feynman-Siegel
gauge and with $\alpha'=1$.
The quadriatic terms in the action involving the transverse photon is
given by
\footnote{
The action in \eq{photon}  agrees with that of 
\cite{kossam} (Eq. (4.7)) till level 2,  with 
$\alpha'=1$, $g=2$, $t=\phi$, $u=-\beta_1$, and $v= B$. }
\be
\label{photon}
S= \int dk A_\mu(-k)A^\mu(k) \left( \frac{k^2}{2} +
e^{-2k^2\ln(3\sqrt{3}/4)} {\cal R}(k) \right)
\ee
Where ${\cal R}(k)$ is given by
\bea
{\cal R}(k) &=& \frac{3\sqrt{3}}{4} t - 
(\frac{49}{12\sqrt{3}} - \frac{4}{3\sqrt{3}} k^2 ) 
\frac{v}{\sqrt{13}} +
\frac{ 11 }{ 12\sqrt{3} } u \\ \nonumber
&+& (\frac{1579}{162\sqrt{3}} - \frac{88}{81\sqrt{3}} k^2 ) A 
+ (-\frac{539}{324\sqrt{3}} + \frac{44}{81\sqrt{3}} k^2 ) F
\\ \nonumber
&+&\frac{20}{27\sqrt{3}}D + \frac{19}{108\sqrt{3}} E
-\frac{20}{81\sqrt{3}}C \\ \nonumber
&+& (\frac{785}{54\sqrt{3}} - \frac{520}{81\sqrt{3}} k^2 +
\frac{64}{81\sqrt{3}} k^4 )B
\eea
Here the variables $t, u, v, A, B , C, D, E, F$ stand for the fields
as defined in \cite{senzwi}.
It is  now easy to extract the kinetic term. It is given by
\be
S_{\rm{KE}} = \int dk {\cal T} A_\mu (-k) k^2 A^\mu (k)
\ee
where
\bea
{\cal T} &=& \frac{1}{2} - 2\ln(\frac{3 \sqrt{3}}{4}) \left(
\frac{3\sqrt{3}}{4} t - \frac{49}{12\sqrt{39}}v +
\frac{11}{12\sqrt{3}} u + \frac{1579}{162\sqrt{3}} A
-\frac{539}{324\sqrt{3}} F  \right. \\ \nonumber
&+& \left. \frac{20}{27\sqrt{3}}D +
\frac{19}{108\sqrt{3}} E - \frac{20}{81\sqrt{3}}C +
\frac{785}{54\sqrt{3}} B \right) \\ \nonumber
&+& \frac{4}{3\sqrt{39}} v - \frac{88}{81\sqrt{3}} A +
\frac{44}{81\sqrt{3}}F - \frac{520}{81\sqrt{3}} B
\eea
We tabulate the rate of decrease of the kinetic term 
of the transverse photon as we approach
the tachyonic condensate by level trucation.
\bea
\begin{array}{ccc}
\hbox{Level  } & \hbox{  Coefficient of   }k^2 & \;\;\; \% 
\hbox{  Decrease} \\
\hline
(1,2) & .1899 & 37.9\% \\
(2,4) & .1861 & 37.2\% \\
(4,8) & .1814 & 36.3\% \\
\hline
\end{array}
\eea
Here the last column shows the decrease in the coefficient of the
kinetic term as compared to $1/2$, which is the value at the
perturbative vacuum. We have used the values found in \cite{senzwi}
for the fields $t, u, v, A, B, C, D, E$ and $F$.
Thus the rate of decrease of the kinetic term is
much slower than  
the rate at which the minimum of the tachyon potential  
is reached in level truncation. In \cite{sen2} it was shown in the  
Dirac-Born-Infeld action the coefficient of the kinetic term is the
tachyon potential. Thus we see clearly that 
the effective action from Witten's open string
field theory is not the Dirac-Born-Infeld action. We have shown that
there is a field redefiniton which relates these two actions. 
This field redefintion is not unique. It will be interesting to
find that unique field redefiniton which relates the two action
explictly.

\subsection{Vector excitations at the tachyon condensate}

To find the physcial excitations of the transverse photon we
evaluate the 
physical poles in the two point function of the transverse
photon. This is done by evaluting the zeros of the funtion
\be
\label{zero}
f(k_0)= -\frac{k_0^2}{2} + e^{2k_0^2}\ln(3\sqrt{3}/4) {\cal R}(k_0)
\ee
Here we have substitued $k^2 = -k_0^2$ in the two point funtion of the
transverse photon from \eq{photon}. Poles in $f(k_0)$ represent masses
of physical excitations.  We find the following results
\bea
\label{polet}
\begin{array}{cc}
\hbox{Level   } & \hbox{  Zeros of the two point function} \\
\hline
(1,2) & \hbox{ No real zeros in  }k_0^2 \\
(2,4) &  k_0^2= 15.6031 \\
(4,8) &  k_0^2 = 13.4106 \\
\hline
\end{array}
\eea
We note that the location of the pole in level 4 decreases as compared
to level 2. To find the physical excitations at the tachyon condensate
it is not only sufficient to look at the poles in the transverse
photon two-point function. We also need to find the zeros in the
determinant of the fluctuation matrix of all the vector particles.
Above level $(2,4)$, vector fields at level 3 mix with the transverse
photon.

We will now show that the pole found at level (2,4) 
exists even when we
consider all the vector fields at level (3,6). Consider the action
including all the vector fields till level 3. We write the 
quadriatic terms in the action as
\bea
S &=& \int dk A_\mu (-k)A^\mu(k)\left (\frac{k^2}{2} +
e^{-2k^2\ln(3\sqrt{3}/4)} {\cal R}(k) \right) \\ \nonumber
&+& 2 A_\mu(-k) \alpha_i(k) V_\mu^i(k) +
V_\mu^i(-k) M_{ij}(k) V_{\mu}^j(k)
\eea
Where $V_\mu^i$ are all the other vector fields till level 3.
These are
\be
\label{fields}
V_\mu^1 \alpha^\mu_{-3} c_1 |0\rangle + V_\mu^2 \alpha_{-1}^{\mu}
b_{-1}
c_{-1} c_1| 0\rangle + V_\mu^3 \alpha_{-1}^{\mu} L_{-2} c_1 |0\rangle
\ee
Evaluating the zeros of the determinant of the fluctation matrix is
equivalent to elimating the fields $V_\mu^i$ using their equations of
motion and then evaulating the zeros of the two point funtion for the
transverse photon. This is given by
\bea
S &=& \int dk A_\mu (-k)A^\mu(k)\left (\frac{k^2}{2} +
e^{-2k^2\ln(3\sqrt{3}/4)} {\cal R}(k) \right. \\ \nonumber
&-& \left. \alpha_i (k) M^{-1}_{ij} (k) \alpha_j (k) \right)
\eea
Now we look at the zeros of the function
\bea
g(k_0) &=& f(k_0) - h(k_0) \\ \nonumber
 &=& f(k_0) - \alpha_i(k_0) M^{-1}_{ij} (k_0) \alpha_j (k_0)
\eea
Here $f(k_0)$ is defined in \eq{zero}. $g(0)$ must be positive, This
represents the (mass)$^2$ of the transverse photon. 
It is positive as there is no tachyonic excitations at the stable
vacuum. For $k_0 \rightarrow\infty$ we have $f(k_0) \rightarrow -v
k_0^2 e^{k_0^2 \ln(3\sqrt{3}/4)}$. We are looking at terms in the
action till level 6. 
The value of $v$ is positive at the tachyon condensate. 
The $f(k_0)$ tends to $-\infty$ as $k_0\rightarrow \infty$.
We examine the behaviour of $h(k_0)$ as $k_0$ tends to infinity.
To do this we look at terms which has the highest power of momentum in
$h(k_0)$. These are given by fields which have the highest power of
$L_{-2}$. For each $L_{-2}$ there is a power of $k^2$. 
Using this we find
$h(k_0) \rightarrow v^2 k_0^4 e^{k_0^2 \ln(3\sqrt{3}/4)}/t$ as
$k_0\rightarrow \infty$. Thus $g(k_0)$ continues to be negative for
large values of $k_0$. As $g(k_0)$ is positive at $k_0=0$, there exists
at least a single zero by continuity. Therefore the pole found at level
$(2,4)$ persists at level $(3,6)$.

We consider the case of level $(4,8)$. Again $g(0)$ is positive as it
represents the (mass)$^2$ of the transverse photon. For
$k_0\rightarrow \infty$, we see that $f(k_0) \rightarrow B k_0^4
e^{k_0^2\ln(3\sqrt{3}/4)}$ and $h(k_0) \rightarrow -B^2k^6_0
e^{k_0^2\ln(3\sqrt{3}/4)}/v$. Thus $g(k_0)$ is positive for large values
of $k_0$. Therefore this analysis is not conclusive. But, it does
suggests that the zero could be removed or might be shifted to a higher
value than given in \eq{polet}. We also note that this method of
analysis of the zeros using momentum dependence simplifies the task of
finding zeros. We need only to look at a subset of correlation
functions.

\section{Conclusions}

We have seen that the naive low energy effective action ${\cal S}
(\ta_\mu , \tilde{\phi})$ does not correspond to the world volume
action of unstable branes. To obtain the world volume action from
string field theory, one has to redefine the tachyon and the gauge
field. We have shown that this is possible to the first non-linear
order in the fields. We have also identified a string field theory
gauge symmetry which corresponds to the $U(1)$ gauge transformation of
the gauge field which appears in the low energy effective action.
These considerations help in understanding the discrepancy of gauge
invariance obtained for certain terms in the non-Abelian Dirac Born
Infeld action of \cite{wati}.
It is easy 
to see that the low energy effective action on unstable branes
contains not only the trivial $U(1)$ gauge symmetry, but a large gauge
symmetry arising from the large gauge symmetry of string field theory.
It would be interesting to understand this symmetry further.

We examined the vector excitations till level $(4,8)$. We showed that
there is a physical vector excitation at level $(3,6)$. At level
$(4,8)$ our methods are not conclusive. But we see that this
excitation can be removed or pushed to a higher energy.

\acknowledgments

The author thanks A. Sen for
useful correspondence and for an initiation into this subject. 
He thanks N. Itzhaki for discussions and
participation at the early stage of this project. 
He acknowledges useful 
discussions with D. Gross, V. Periwal and M. Srednicki.
He thanks the participants of the Santa Barbara  post-doc journal
club for discussions. 
The work of the author is supported by NSF grant
PHY97-22022.

\end{document}